# SYNTHESES AND EVALUATION OF $Fe^{2+}$, $Co^{2+}$ AND $Ni^{2+}$ COMPLEXES WITH DERIVATIVES OF VANILLIN – TRYPTOPHAN SCHIFF BASE LIGAND

SALISU ABUBAKAR [1], Prof. G. A. Shallangwa* [2] and Dr. Abdulkadir Ibrahim[2]

[1]Department of Chemistry Federal College of Education, P.M.B. 1041, Zaria
[2] Department of Chemistry, Ahmadu Bello University, Zaria, Nigeria.

*Corresponding Author's abubakarsalisu@abu.edu.ng +2348136135156

**Abstract**

This research focused on synthesis, characterization and determination of biological activities of $Fe^{2+}$ $Co^{2+}$ and $Ni^{2+}$ complexes of a novel *L*- Tryptophan-Vanillin schiff base ligand. They were synthesized and characterized by determination of their Purity, Solubility, Elemental analyses using phase match XRD data, FT-IR spectra, Molar conductivities, Magnetic susceptibilities, PXRD and Biological activities. Results from the analyses revealed that, mpt of schiff base ligand (SL) = 84 – 85 °C, $Fe^{2+}$-SL = 245 - 246 °C, $Co^{2+}$-SL= 271 - 272 °C while $Ni^{2+}$ - SL is >350 °C. Molar conductivities were found to be 10300, 5000, 17300 and 52900 $Sm^2\,mol^{-1}$ for the schiff base and the complexes respectively. It indicates that, the complexes are non-electrolytic in nature. Magnetic susceptibilities of the complexes found to be higher than the spin only values due to unpaired electrons ($Fe^{2+}$ = 5.73 B.M, $Co^{2+}$= 4.52 B.M while for $Ni^{2+}$ = 3.46 B.M, suggesting an octahedral geometries. Electronic absorption ($\lambda_{max}$) for $Fe^{2+}$ = 480, $Co^{2+}$ = 520nm while $Ni^{2+}$ shows two bands at 480 and 570 which signifies, n-$\pi^*$ and $\pi$-$\pi^*$ transition respectively. Their crystalinity index was also assessed using pxrd technique, it shows that complexes are 75, 80 and 85 % crystalline with average crystallite size of 21.63 nm. Antimicrobial test results revealed that, $Co^{2+}$ and $Fe^{2+}$ complexes have excellent activities against gram negative bacteria (E. coli, F. Shigella . and S. Typi). They all shows efficient activities against some gram positive bacteria such as: *MRSA*, *Klebsiella P*. and *S. Pnemoniae and* some fungi spp *A. niger* and *C. albicans*. Therefore based on the findings from the study it was concluded that, schiff base is bidentate ligand and octahedral geometries, paramagnetic susceptibilities were suggested for the complexes.

# 1.0 Introduction

Schiff bases from amino acids and flavones such as vanillin and quercetin and their complexes have numerous biological activities (Akhter *et al.,* 2017b; Sani and Hussain, 2017). The high affinity for the chelation of the schiff bases towards the transition metals is utilized in preparing their solid complexes and due to its polyhydroxylated chemical structure, vanillin easily forms complexes with species having free orbitals which may be occupied (Barbosa *et al.,* 2019).

The potentials of vanillin schiff base ligands in inorganic chemistry can be understood very well from the properties exhibited by its complexes such as: magnetics; luminescence, chirality, catalysis, cytotoxicity and ferro electricity among others (Al-jeboori and Al-shimiesawi, 2021; Neacşu *et al.,* 2018 and Barbosa *et - al.,* 2019). Vanillin derivatives are found to be active against both gram-positive and gram-negative bacterial strains and has been shown to be effective against some fungal spp ( Sheyin *et al.,* 2018).

*L*-Tryptophan (Trp) is an essential amino acid which is required for the biosynthesis of proteins in the body. It is very significant for nitrogen balance, maintenance of muscle mass and body weight in humans (Yan *et al.*, 2010).

## 1.1 The Chemistry of Schiff base Formation

In reactions that yield schiff base ligands, the electrophilic carbon atom of carbonyl compounds will experience a nucleophilic attack by amines to give a compound with C=O being replaced by C=N. The reaction is reversible and occurs as shown in eqn (1) or (2) depending whether the carbonyl is an aldehyde or a ketone ( Usman *et al.,* )

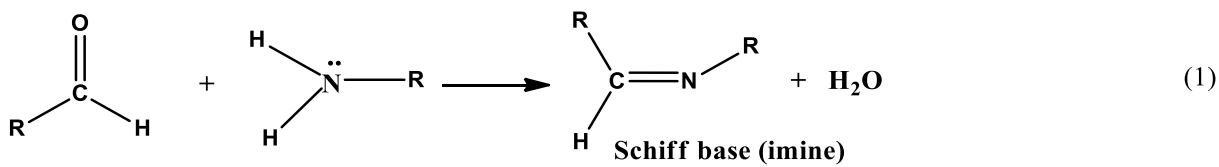

(1)

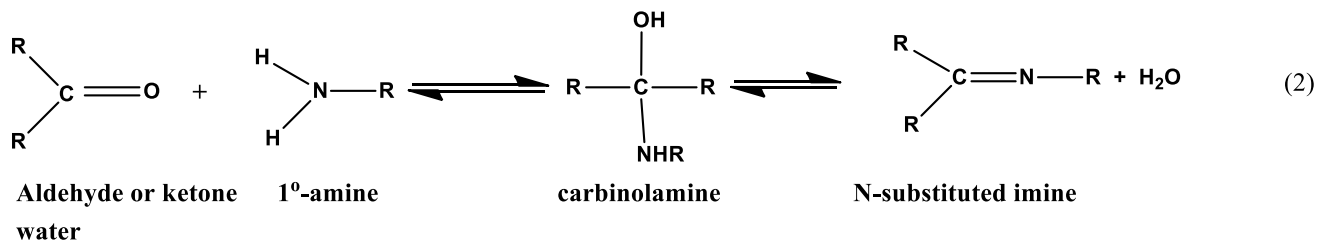

(2)

**Aldehyde or ketone**  **1°-amine**  **carbinolamine**  **N-substituted imine**
water

Condensation products of vanillin with amines confers biological activities as well as having good complexation ability with metal ions (Saranraj and Devi, 2018). This study focused on synthesis, characterization and test for antimicrobial activities of a schiff base ligand derived from condensation of *o*-vanillin with *L*– tryptophan and complexes of their ligands with Fe2+, Co 2+ and Ni2+ ions. Transition metal complexes with bidentate schiff base ligands that contains nitrogen, oxygen, or sulphur donor atoms contribute immensely in biological systems (Malik *et al.*, 2018; Maalik *et al.,* 2014).

Numerous researches carried out on transition metal complexes derived from schiff base ligands have pointed out their significance due to their properties and wider biological applications (Nagesh and Mruthyunjayaswamy, 2015; Agbese *et al*, 2018). Fugu *et al.* (2013) reported that schiff base complexes derived from 4-hydroxysalicylaldehyde and amines have strong anticancer activity. Schiff bases of ketones, their derivatives, and their metal complexes are active anticancer and antioxidant agents (Dhanaraj and Johnson, 2014).

Transition metal ions are amongst the important species needed to keep human body healthy and active because several important biological functions depend upon their presence and their absence may lead to deficiency diseases (Arvind and Andreas 2018). Mustafa and Alsharif (2018) opined that, most notable metals that exist in human body are in the form of ions such as: $Fe^{2+}$, $Co^{2+}$, $Ni^{2+}$, $Ca^{2+}$, $Cu^{2+}$, $Zn^{2+}$, and $Cr^{2+}$. Transition metal complexes are observed to be generally more biologically active than the uncomplexed ligands, they are formed due to the presence of the metal moieties (Mohamed and Omar, 2006; Abu Bakar *et al.*, 2010a).

**2.0 Experimental**

All chemicals used for this research are of analytical grade (AR) obtained from Sigma Aldrich and other reliable chemical vendors. The chemicals were used as purchased.

**2.1 Materials**

The following materials were used for the study: Mettler Analytical weighing balance (MT 200B 0.0001-200G), spectrophotometer (JENWAY-7315), magnetic stirrer (Stuart), fume cupboard (Biobase), melting point machine (SMP 10 Fascia, Stuart Bibby Scientific), FT-IR spectrophotometer (Nicolet iS5-Thermo scientific), electrical conductivity meter (Hanna Instrument, HI 9835, Woon socket R. I USA), heating/drying oven (TT-9053, Techmel and Techmel U.S.A, YLD 2000) and heating mantle among others.

**2.2 Methodology**

**2.2.1 Syntheses of Schiff base Ligand**

The method used to synthesize the ligand was adopted from that of Kaczmarek *et al.* (2009) with minor modifications. Solution of vanillin (0.30g, 2mmol) was mixed with solution of *l*-tryptophan (0.4g, 2mmol) in 15 ml ethanol with 1ml dilute sodium hydroxide solution (0.056g, 1mmol) solution. The reaction mixture was stirred for about 10 minutes and refluxed over a water bath for 5hrs at 60 °C. The coloured precipitate formed was filtered, wash with 50% cold ethanol and rinse with diethyl ether. It was allowed to dry in a vacuum desiccator to constant weight which

yield = 72.0% schiff base ligand (SL). The yellowish product was recrystallize in methanol for further purification.

## 2.2.3 Syntheses of the Complexes

The complexes were prepared using standard methods of (metal: ligands ratio) adopted from Haddad (2016); Asgharpour *et al.,* (2017) and Hossain *et al.* (2018) with some minor modifications. An ethanolic solution of schiff base (20 mmol) was prepared and added drop wise to the solutions of metal (II) chloride (10.0 mmol) $FeCl_2$, $CoCl_2$ and $NiCl_2$ stirred for about 20 minutes. Same concentrations were prepared and refluxed in water bath at 60°C for 6 hours. The pH of the mixture was adjusted to about 7.5 with few drops of potassium hydroxide solution and the coloured solid formed were filtered, wash with 50% cold ethanol, and finally dried in a vacuum desiccator to constant weight. The synthetic routes is summarised in equations (3) and (4):

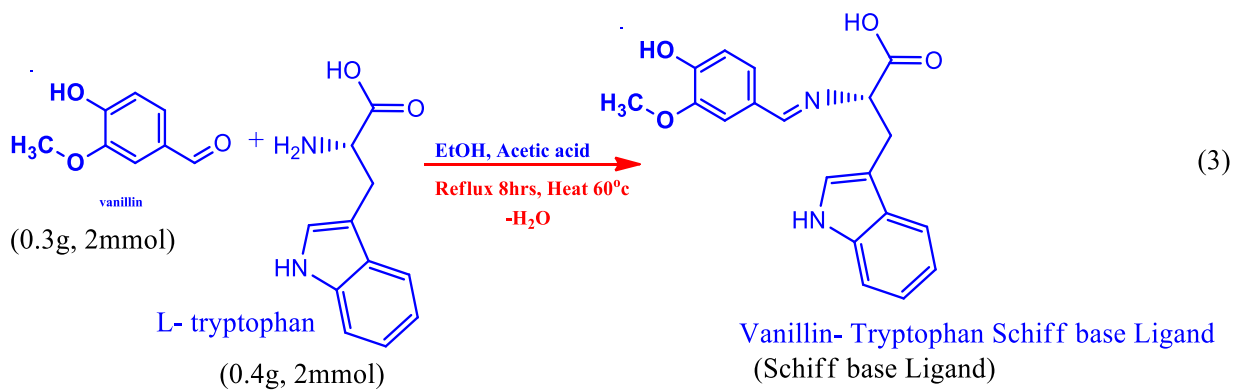

(3)

Fig. 1: A flowchart of Syntheses of Schiff base Ligand and the Metal (II) Complexes

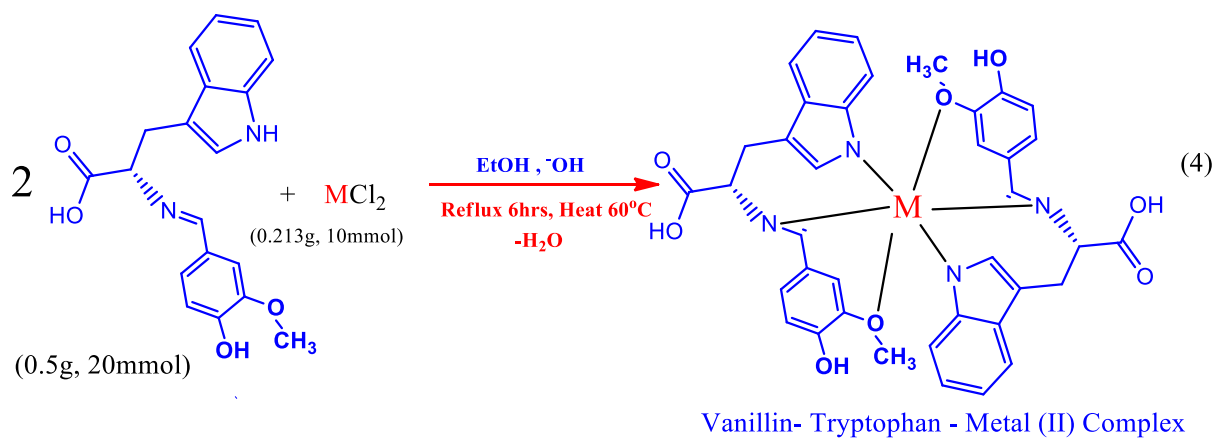

Fig. 2: A flowchart of Syntheses of Metal (II) Complexes

**2.4 Physical Measurements**

Solubilities of the schiff base and metal complexes were checked in water, 50% ethanol and DMSO. Melting points of the complexes were also determined to established their purity using electrically heating melting point apparatus (SMP10, Fascia, Stuart; Bibby scientific), molar conductivity were assessed to test the electrolytic nature of the complexes (with an Hann scientific Electrical conductivity meter). The fourier transform infrared spectra were determined with FT-IR spectrometer (Nicolet iS5, Thermo scientific), wave length of maximum absorption ($\lambda_{max}$) of the schiff base and the complexes were measured using an aqueous solution of the compounds in uv/visible spectrophotometer (Jenway, 7315) 1cm cuvette at (200 – 780 nm). Elemental analyses was performed using the match phase crystals analysis software and the report obtained from the PXRD data. Magnetic susceptibility followed by x-rays diffraction technique were used to determine geometries, degree of crystalinity, and the morphological properties of the complexes.

**2.5 Assessment of Antimicrobial Activities**

**(i) Minimum Inhibitory Concentration (MIC)**

The MIC values was determined in accordance with standard method proposed by clinical and laboratory standard institute (CLSI, 2019) using a broth dilution techniques. Mc Farlands' turbidity standard was prepared and used for the study. 1ml of sterilized MH- media were added to each of clean, dried and sterilized test-tubes, followed by 1mL of 0.5 MCF bacterial suspension which contained about ($1-2\times10^8$ CFU/mL bacterial load). 1mL of 2000 μg / mL solution of the test compounds was transferred into each test tube and serially diluted to give concentrations of (1000, 500, 250, 125, 62.5, 31.25, 15.625 μg/mL).

The solution was added to the mixture followed by 0.1 mL of the bacterial suspension in a normal saline solution. The test tubes were incubated at 37 °C for 24 hrs to enhance bacterial growth, test-tubes with lowest and the highest turbidity were examined, compared with standard and the results were carefully recorded.

**(ii) Minimum Bactericidal Concentration (MBC)**

The MBC was assessed using the method developed by Hadariah *et al.* (2019) to ascertain whether the test pathogens were completely killed or just inhibited their growth. Mueller Hinton agar was sterilized, poured into petri dishes and the media was allowed to cool/solidify. An aliquots from samples which shows sensitive to the MIC test were carefully selected and subjected for MBC evaluation. They were sub-cultured and incubated at 28° C for 7 and 14 days in some cases. To determined CFU, the plates of the media was taken for microscopy and observed for bacterial growth.

**(iii) Minimum Fungicidal Concentration (MFC)**

Sarah *et al.* (2016) and CLSI (2019) methods was adopted to carry out this test using micro broth dilution technique with little modifications. Two fungal spp (*Aspergillus Niger* and *Candida albicans*) isolates were used, grown for 7 days at 28 °C incubation period in potato dextrose agar. The stock suspension was diluted with normal saline solution to corresponding concentration of $1\times10^5$ cfu /mL and viability of the fungi spp was confirmed using the sabouraud dextrose agar plates and counting the number of cfu/mL

Two fold serially diluted solution of the test compounds were prepared to obtained 200μg/ml, 100μg/ml, 50μg/ml, 25μg/ml and 12.5μg/ml respectively. A micro pipette was used to transfer 0.1ml of the test fungal strain solution into the test-tubes containing the mixture with varied concentrations of the compounds. The mixture was incubated at 38 °C for 72 hours after which the turbidity was observed and compared with standard for the growth of the pathogens. The turbidity (growth) between the test-tube which contained lowest and the highest concentration of the compound were examined and compared with standard. The results were carefully recorded in accordance with clinical and laboratory standard institute CLSI M100, (2019).

## 3.1 Results and Discussion

**Table 1: Shows some Physical Parameters test Results for Schiff base and Metal Complexes**

| Products | Colour | %Yield | Solubility | Mpt (°C) | $\Lambda_m$ (Sm² mol⁻¹) | µeff (B.M) |
|---|---|---|---|---|---|---|
| Vanillin – Tryptophan SBL | Golden Yellow | 72.0 | SS ($H_2O$) SL ( EtOH) SL DMSO) | 84 | 10300 | - |
| Vanillin – Tryptophan - $Fe^{2+}$ | Greenish yellow | 85.0 | SL ($H_2O$) SS (EtOH) SL ($H_2O$) SS (EtOH) | 245 | 5000 | 5.73 |
| Vanillin – Tryptophan - $Co^{2+}$ | Brownish yellow | 81.0 | SL ($H_2O$) SL(EtOH) SL(DMSO) | 271 | 17300 | 4.52 |
| Vanillin – Tryptophan - $Ni^{2+}$ | Yellowish-green, | 59.4 | SS ($H_2O$) SL (EtOH) SL(DMSO) | >350 | 52900 | 3.46 |

*Keys: SL = soluble, SS = sparingly soluble*

The results in Table 1, indicates colour of the schiff base and the metal complexes, percentage yields and solubilities in some solvents. It revealed that, schiff base ligand is sparingly soluble in water, but it readily dissolved in DMSO and 50% ethanol solutions solution while the complexes are soluble in water, DMSO respectively. It also indicated that, schiff base has lower melting point (84 °C) than the complexes, but Ni2+ complex has the highest melting temperature of (> 350 °C) followed by Cobalt (II) complex (271 °C) while Iron (II) complex has (245 °C).

Molar conductivity measurement shows that, both the ligand and the complexes were non electrolytic in nature and magnetic susceptibility measurement obtained at room temperature revealed that, complexes are paramagnetic with observed magnetic moment as: $Fe^{2+}$ = 5.73 B.M, $Co^{2+}$ = 4.52 B.M while for $Ni^{2+}$ = 3.46 B.M respectively which are slightly higher than the spin only values due to unpaired electrons suggesting an octahedral geometry for all the complexes. This is similar to the work of Zurowska and Bauskey (2011).

**Table 2: UV-visible Data for the Schiff base Ligand and the Complexes**

| S/N0. | Compounds | Wave length λmax (nm) | Transition Assigned | Geometries |
|---|---|---|---|---|
| 1. | Schiff base Ligand (SL) | 255 – 350 | $\pi - \pi^*$ (C =C in benzene) | ---- |
|  |  | 330 – 380 | $n - \pi^*$ (C = N in imine) |  |
| 2. | SL – Fe2+ Complex | 460 – 480 | $\pi - \pi^*$ (C =C in benzene) | Octahedral |
| 3. | SL – Co 2+ Complex | 500 – 520 | $n - \pi^*$ (C = N in imine) | Octahedral |
| 4. | SL – Ni2+ Complex | 550 – 570 | $\pi - \pi^*$ (C=N in imine) | Octahedral |
|  |  | 369 – 480 | $\pi - \pi^*$ (C=C in benzene) |  |

Results in Table 2, shows electronic transitions within the schiff base and the complexes scanned from 300 – 750 nm in ethanolic solutions (Fig. 2-4) which indicates some bands as $n - \pi^*$ and $\pi - \pi^*$ transitions from to non-bonding to anti bonding molecular orbitals and delocalization of pi-electrons within the benzene rings. The absorption bands assigned and the proposed geometries of the complexes are as follows: $Fe^{2+}$ 460 – 480 nm, $Co^{2+}$ shows only one bands at 500 – 520nm while $Ni^{2+}$ shows two absorption bands at 369 – 480 and 550 –570nm respectively. This result is in agreement with that of Suesh and Pracash (2010) and (Crystal, Y. et –al., 2015).

**Table 3: Elemental Analysis of the Ligand and the Complexes**

| Products | % of Elements Found (Calculated) | | | | | | | Formulae |
|---|---|---|---|---|---|---|---|---|
|  | C | H | O | N | Fe | Co | Ni |  |
| SBL | 52.35 (52.79) | 5.12 (4.63) | 17.73 (16.90) | 24.8 (24.01) | - | - | - | $C_{19}H_{18}N_2O_4$ |
| SBL – $Fe^{2+}$ | 39.34 (39.72) | 4.18 (3.96) | 23.03 (22.89) | 3.37 (2.65) | 26.88 (26.14) | - | - | $C_{38}H_{36}N_4O_8Fe$ |
| SBL – $Co^{2+}$ | 47.04 (46.31) | 9.71 (8.99) | 10.36 (10.20) | 12.01 (11.98) | - | 20.88 (20.09) | - | $C_{38}H_{36}N_4O_8Co$ |
| SBL – $Ni^{2+}$ | 45.23 (45.18) | 4.74 (4.98) | 25.11 (25.01) | 4.40 (4.85) | - | - | 20.52 (20.90) | $C_{38}H_{36}N_4O_8Ni$ |

Table 3 shows result of elemental analysis and confirmed molecular formula of the schiff base and the metal complexes, it revealed that, the complexes are mononuclear with 2moles of the ligand bonded to 1mole of the metal ion. This implies that, the metal – Ligand ration is 2:1 with molecular formula as [M-(SBL)$_2$. H$_2$O] where SBL = Schiff base ligand, M = Co$^{2+}$ and Ni$^{+2+}$ respectively.

**Table 4: FT-IR Spectral mode (350 – 4000 cm$^{-1}$) of the Schiff base and the Metal complexes**

| Complexes | V(HC=N) | V(M—N) | V(M—O) | V(COO) | V(C—O) | V(O—H) | V(C—N) | V(C-O-C) | V(C=C) |
|---|---|---|---|---|---|---|---|---|---|
| SBL | 1663 sh | - | - | 1362 sh | 1282 sh | 3326 m | 1373 sh | 1625 m | 1521 sh |
| SBL – Fe$^{2+}$ | 1583.36 sh | 401.64 sh | 535.30 sh | 535.30 br | 1026.32 sh | 3208.51 br | 1509.46 sh | - | - |
| SBL– Co$^{2+}$ | 1583.45 sh | 463.69 sh | 511.36 sh | 588.73 sh | 1025.14 | 3211.86 br | 1506.96 sh | - | - |
| SBL – Ni$^{2+}$ | 1586.13 sh | 419.78 sh | 385.20 sh | 410.36 sh | 1017.85 sh | 3362.07 br | 1630.99 sh | - | - |

*Keys: SBL = Vanillin – Tryptophan Schiff base Ligand,*
*SBL – Fe$^{2+}$= Vanillin – Tryptophan Schiff base – Fe2+ complex, SBL– Co$^{2+}$ = Vanillin – Tryptophan Schiff base – Co 2+ complex, SBL – Ni$^{2+}$ = Vanillin – Tryptophan Schiff base – Ni$^{2+}$ complex*

Results in Table 4 above, shows FT-IR spectra bands of the ligand and the complexes at 1628 v (HCN, str), 382.47 v(O-H str) and 362.12 cm$^{-1}$ respectively. The FT-IR analysis revealed that, the bonding of the schiff base to the metal centres has occur as expected. In the schiff base ligand, a strong sharp band observed at 1663 cm$^{-1}$ can be assigned to V (HC=N) azomethine stretching vibration. When reacted to form complex with the metal ions, the band shifted to lower frequencies in 401.64 - 419.78 cm$^{-1}$ range for v(M—N) and 385.2 - 535.30 cm$^{-1}$ range for v(M—O) indicates that, new bonds has formed that, azomethine coordinated to the metal ions through Oxygen and Nitrogen donor atoms in the schiff base ligand.

The symmetric carbonyl stretching v(COO$^-$) is also shifted to a higher frequency from 1362cm$^{-1}$ to 535.30, 588.73 and 410.36cm$^{-1}$ region for Fe$^{2+}$, Co$^{2+}$ and Ni$^{2+}$ complexes respectively. Similarly the v (C – O) phenolic band in the complexes shows a significant shift to a lower frequencies of 1282 cm$^{-1}$ in the ligand, 1026.32, 1025.14 and 1017.85 cm$^{-1}$ for the metal ions respectively. The spectral of the complexes shows broad bands between 3208.51, 3211.86 and 3362.07 cm$^{-1}$ range which attributed to O–H stretching vibration from phenolic group (Fig.5,6and7). Information obtained from the FT-IR results of some of the major functional groups in the schiff base and the complexes revealed that, the schiff base is bidentate ligand and it bound to the metal ions through the phenolic oxygen and azomethine nitrogen atoms to form an octahedral complexes.

**Table 5: Powder X-rays Diffractions Analysis**

| Complexes | Peak position 2θ (º) | FWHM β (º) | % Crystalinity Index (CI) | Crystallite size D (nm) | Average D (nm) |
|---|---|---|---|---|---|
| Fe$^{2+}$– SL | 53.40 | 0.654 | 75 | 19.41 | |
| Co$^{2+}$ - SL | 58.62 | 0.678 | 80 | 21.32 | 21.63 |
| Ni+$^{2+}$ – SL | 62.45 | 0.545 | 84 | 22.16 | |
| | K = 0.94, | λ = 0.5418 (nm) | | | |

Table 5 shows results of PXRD pattern of the complexes scanned at laboratory temperature (303 K) with an XRD machine (RIGAKU). The degree of crystalinity and crystallite sizes and average crystallite size of the complexes were obtained within the range of 2θ values from 10 – 120º at wavelength of $^\lambda$CuK$_{α1}$ radiation = 1.5418 nm. The XRD results further supported by the inter planar distance and the crystallite sizes (d) of 19.41, 21.32 and 22.16 corresponding to Fe$^{2+}$– SL, Co$^{2+}$ - SL and Ni+$^{2+}$ – SL respectively.

**Table 6: Anti-microbial and Antifungal Sensitivity Test for the Schiff base and Complexes**

| S/N0. | Products | MRSA | S. Pnemoniae | E. Klebsiella Pneumoniae | ESBL E. Coli | F.Shigella | S. Typi | C. Alicans | A. Nigger |
|---|---|---|---|---|---|---|---|---|---|
| 1. | SBL | R | R | MS | MS | MS | R | MS | MS |
| 2. | SBL – $Fe^{2+}$ | R | R | S | R | S | R | S | R |
| 3. | SBL – $Co^{2+}$ | MS | S | S | S | R | R | MS | S |
| 4. | SBL – $Ni^{2+}$ | MS | S | S | S | R | MS | R | S |
| 6. | Standard 1* | - | - | - | MS | - | MS | MS | - |
| 7. | Standard 2* | - | - | - | - | - | - | MS | - |
| 8. | Control | - | - | - | - | - | - | - | - |

**Key: S** = sensitive, **MS** = mild sensitive, **R** = resistance, **-** = negative (no reaction), **C** = control, Standard 1* = Augmentin, Standard $_2$* = Terbinafine

Table 5 shows sensitivity test results of the products against some selected pathogens used for the study. It indicates that, one gram +ve with two gram –ve bacteria and the fungal spp shows mild sensitive reactions to the schiff base ligand, *Klebsiela P., Shigella* and a fungi sp (*Candida A.*) are very sensitive to $Fe^{2+}$ complex, $Co^{2+}$ complex shows its effectiveness to *streptococcus P., Klebsiela P.,* E. coli and *Aspergillus N.* While $Ni^{2+}$ complex revealed to be very effective and positive sensitivity test to *streptococcus, Klebsiela, E. coli and Aspergillus N. fungal sp.*

**Table 7: Shows Zone of Inhibition (Z.I) Test Results of the Schiff base Ligands and their Metal (II) Complexes in (mm)**

| S/N0. | Compound | MRSA | S. Pnemoniae | E. Klebsiella Pneumoniae | ESBL E. Coli | Shigella spp | Salmonella spp | C. Albicans | A. Nigger |
|---|---|---|---|---|---|---|---|---|---|
| 1. | SBL | - | - | 5 ± 0.2 | 7 ± 0.2 | 10 ± 0.4 | - | 8 ± 0.3 | 7 ± 0.1 |
| 2. | SBL – $Fe^{2+}$ | - | - | 22 ± 0.1 | - | 25 ± 0.3 | - | 12 ± 0.2 | - |
| 3. | SBL – $Co^{2+}$ | 20.0 | 23 ± 0.5 | 26.0 | 18.0 | - | - | 20 ± 0.2 | 10 ± 0.5 |
| 4. | SBL – $Ni^{2+}$ | 21 ± 0.5 | - | - | 23 ± 0.5 | - | 13.0 | 23 ± 0.2 | 16 ± 0.4 |
| 5. | Control | - | - | - | - | - | - | - | - |
| 6. | Standard $_1$* | 11 | - | 15 | 13 | - | 17 | - | - |
| 7. | Standard $_2$* | - | - | - | - | - | - | 8.5 | 10 |

**Key: S** = sensitive, **MS** = mild sensitive, **R** = resistance, **-** = negative (no reaction), **C** = control, Standard $_1$* = Augmentin, Standard $_2$* = Tarbinafin

The data in Table 6 above, shows results for zone of inhibition test of the products against some those pathogens that shows positive sensitivity (MIC) test from Table 5. It indicates similarly the regions (in mm) where those pathogens are being killed or their growth are inhibited by the compounds as appropriate where the schiff base shows least inhibition zone of (5, 7, 10, 8 and 7mm) followed by Fe(II) complex (12, 22 and 25 mm), Ni2+ (13, 16, 21, and 23 mm) while Co 2+ complex shows the widest inhibition zone (10, 18, 20, 23, and 26 mm) respectively. The control agar plate shows negative test result with less than 0. 1 mm. this indicates that, all the microbes grow freely in the absence of the compounds.

**Table 8: Shows Results of minimum Bactericidal (MBC) and Fungicidal Concentrations (MFC) (µg/mL) for the Schiff base, metal complexes and Standard Drugs against the Pathogens used for the study**

| S/N0. | Products | MRSA | S. Pnemoniae | E. Klebsiella Pneumoniae | ESBL E. Coli | F.Shigella | S. Typi | C.Albicans | A. Nigger |
|---|---|---|---|---|---|---|---|---|---|
| 1. | SBL | ++ | +++ | -- | - | -- | +++ | -- | -- |
| 2. | SBL – $Fe^{2+}$ | ++ | ++ | --- | ++ | --- | +++ | -- | + |
| 3. | SBL – $Co^{2+}$ | --- | -- | -- | - | + | + | --- | -- |
| 4. | SBL – $Ni^{2+}$ | --- | ++ | + | -- | ++ | -- | -- | -- |
| 5. | Standard$_1$ (mg/mL) | +++ | ---- | ---- | ---- | +++ | ---- | + | +++ |
| 6. | Standard$_2$ (mg/mL) | + | +++ | + | +++ | + | ++ | ---- | + |
| 7. | C | +++ | +++ | +++ | +++ | +++ | +++ | ++ | +++ |

*Keys: ++++ = very intense microbial growth +++ = high microbial growth (100 -250 mg/ml), ++ = moderate microbial growth, ++ = mild microbial growth, - No growth slightly clear at (215 µg/ml), -- No growth very clear at (250 µg/ml), --- No growth, very –very clear at (500 µg/ml), ---- No growth immediate sharp clear at (1000 µg/ml), C = control, Standard$_1$\* = Augmentin, Standard $_2$\* = Terbinafine*

Table 7 shows MBC and MFC results of the test compounds, from the preliminary in vitro antimicrobial activities of the ligand and the complexes. They were screened for their minimum inhibitions concentrations and in vitro antimicrobial actions against the selected microbes (Table 7). The minimum active concentrations measured in (µg/mL) and two standard were used (Augmentin and Terbinafine for fugal spp). The results revealed that, the complexes showed more significantly active for killing the pathogens than the ligand and the overall activities of all the three complexes were found to be very effectives even at very low concentration of less than (250 µg/mL) is being considered to be the minimum concentrations of the schiff base and complexes required to kill (99.99%) of the pathogens completely.

## 3.2 Summary and Conclusion

### 3.2.1 Elemental Analysis

The result of elemental analysis confirmed the proposed molecular formula of the schiff base and the metal complexes ($C_{19}H_{18}N_2O_4$ and $MC_{38}H_{36}N_4O_8$) and it revealed that, the complexes are mononuclear with 2moles of the ligand bonded to 1mole of the metal ion. This implies that, the metal – Ligand ration is 2:1 with molecular formula as [M-(SL)$_2$ .H$_2$O] where SBL = Schiff base ligand, M = $Co^{2+}$ and $Ni^{+2+}$ respectively.

### 3.2.2 Magnetic Susceptibility Measurement

Information regarding geometries of the complexes were obtained from their electronic spectral and magnetic moments and the observed magnetic moment are: $Fe^{2+}$ = 5.73 B.M, $Co^{2+}$ = 4.52 B.M while for $Ni^{2+}$ = 3.46 B.M respectively which are slightly higher than the spin only values due to unpaired electrons suggesting an octahedral geometries for all the complexes.

### 3.2.3 Conductivity measurement

Molar conductivity of the complex in $10^{-3}$ moldm$^{-3}$ DMSO solution were measured at laboratory temperature. The values indicated that, the complexes are non-electrolytic in nature. The result is in line with the findings of Zahid H., and Hazoor A. Shed (2018) and Amadou G. *et- al,* (2018).

### 3.2.4 Spectral Analysis

The electronic spectra of the ligand and the complex shows there bands 1628 v (HCN, str), 382.47 v (O-H str) and 362.12 cm$^{-1}$ respectively. The FT-IR analysis revealed that, the binding of the schiff base to the metal centres has been obtained by comparing the FT-IR –spectra of the ligand with those of the complexes.

### 3.2.5 Electronic Spectrum

The electronic spectrum of the schiff base (Fig. 2 -4) above, shows a broad band at and their metal complexes were determined from the scanned range of 350 – to – 750 visible region using ethanol as solvent. The absorption regions, bands assigned and the proposed geometries of the complexes are shown in Table 2. The scanned electronic transition for $Fe^{2+}$ shows two bands at 460 – 480 nm, $Co^{2+}$ shows only one bands at 500 – 520nm while $Ni^{2+}$ shows two absorption bands at 369 – 480 and 550 –570nm respectively.

### 3.2.6: Powder X-rays Diffractions Analysis

Degree of crystalinity, crystal size of the complexes were obtained and PXRD pattern scanned in the range of 2θ values from 10 – 120º at wavelength $^\lambda CuK_{α1}$ radiation = 1.5418 Å. The results revealed that, the complexes are crystalline with high crystalinity index of 72, 80 and 84% respectively.

### 3.2.7 Uv-visible Spectrum of the Schiff base Ligand and the Complexes

The electronic spectrum of the schiff base and the complexes shows broad bands, when scanned from 350 – to – 750 visible region in ethanolic solution as are shown in Table 1.0 above.

### 3.2.8 Assessment of Biological Activities of the Schiff base and the Metal (II) Complexes

The antibacterial activities test were carryout to find the efficacy of the novel schiff base and the complexes. It was done according to standard procedures in vitro using some clinical strains of microbes *MRSA*, *ESBL Klebsiella Pneumoniae*, *Streptococcus Pnemoniae* , ESBL E. coli, Shigella spp. and Salmonella sp together with two fungi spp *Aspergillus* and *Candida albicans.* The test compounds shows ability to inhibit growth of the microbes which was compared with some antibiotics (Augmentin and Terbinafine). The results revealed that, SL – $Ni^{2+}$ complex shows highest activity followed by SL-$Co^{2+}$ and the schiff base respectively. Comparatively, the inhibition zones of the SL and the complexes at the different concentrations indicated that, the complexes has better activities than the schiff base ligand used for this study.

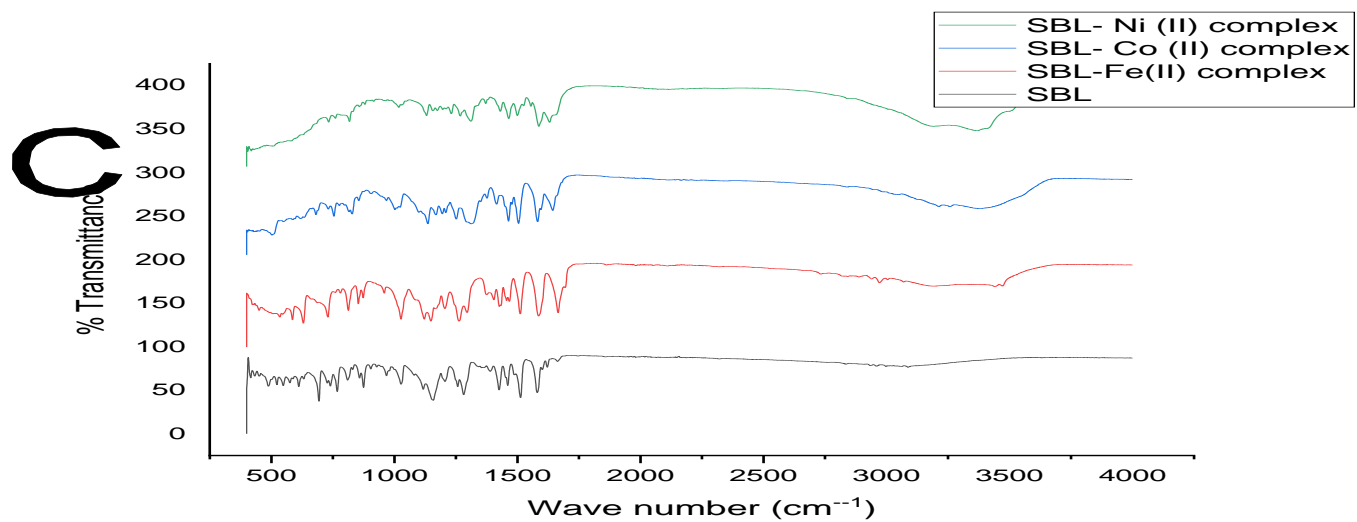

Fig1. FT-IR Spectrum of the Schiff base and the Metal (II) Complexes

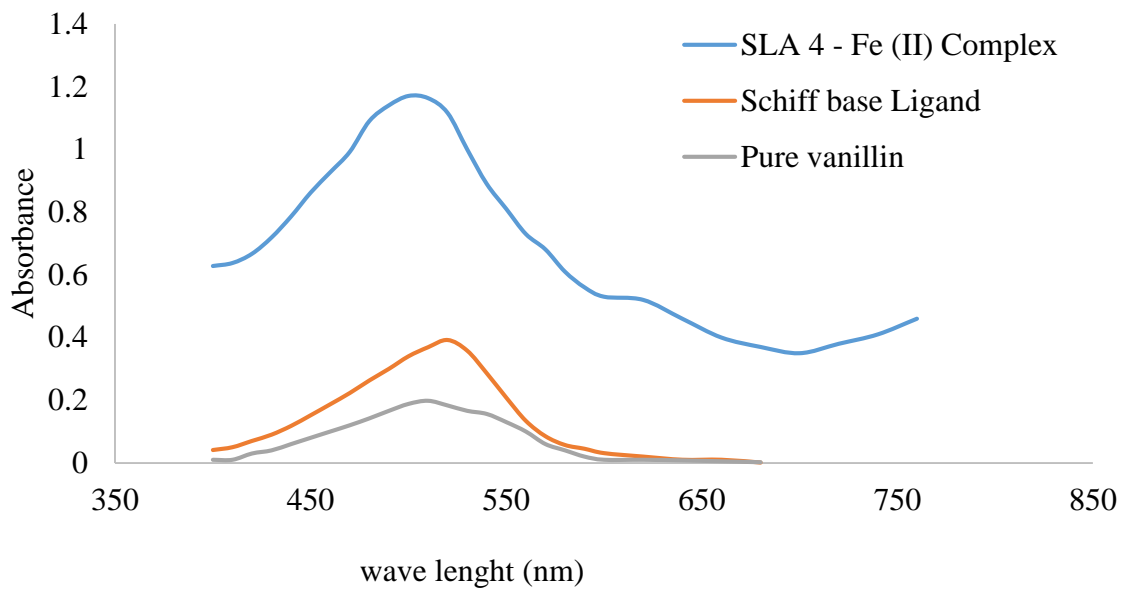

Fig 2: A Staked Absorption Spectra of Pure vanillin, SL and SL – Fe2+ Complex

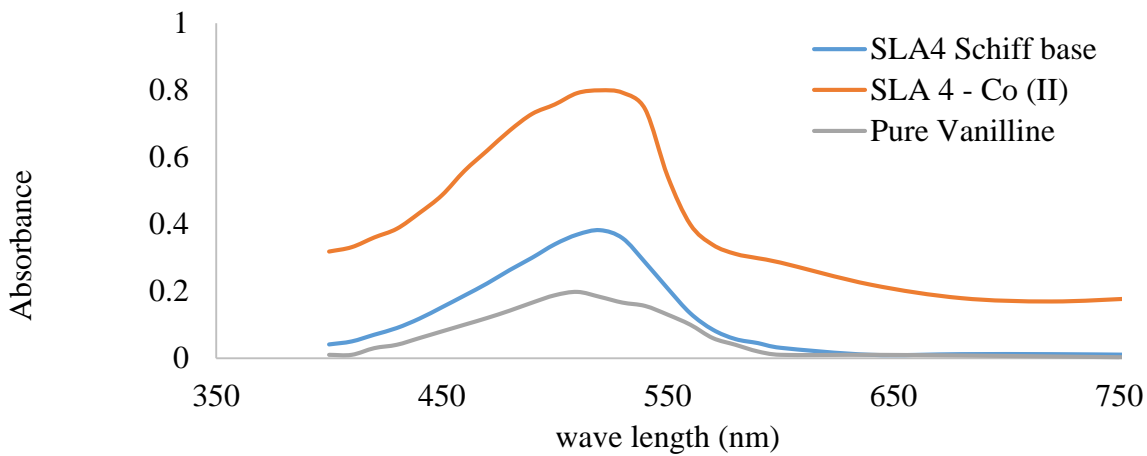

Fig 3: A Staked Absorption Spectra of Pure vanillin, SL and SL – Co 2+ Complex

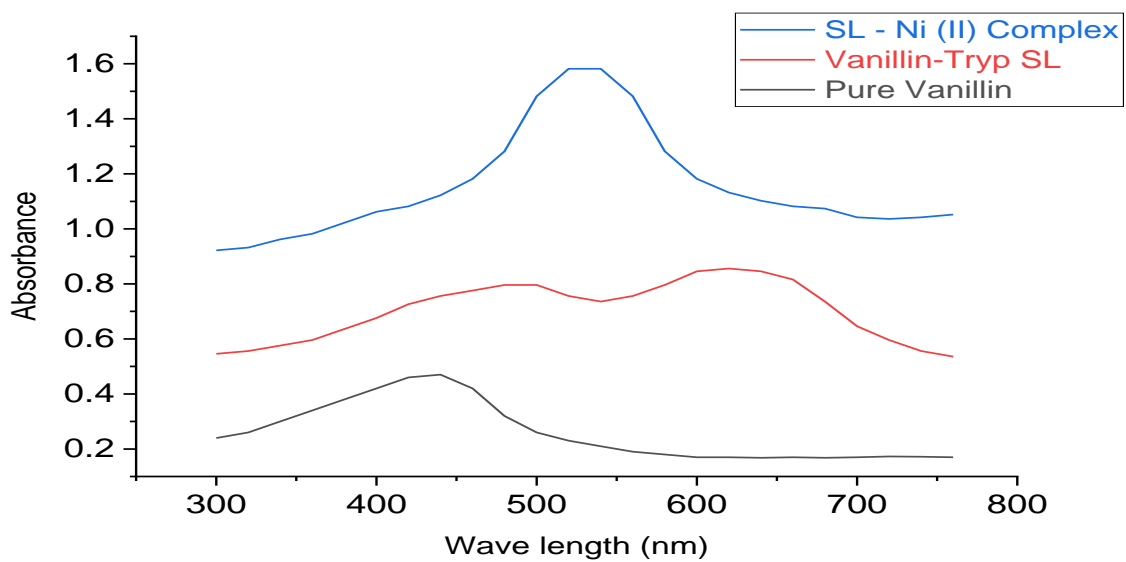

Fig 4: A Staked Absorption Spectra of Pure vanillin, SL and SL – Ni2+ Complex

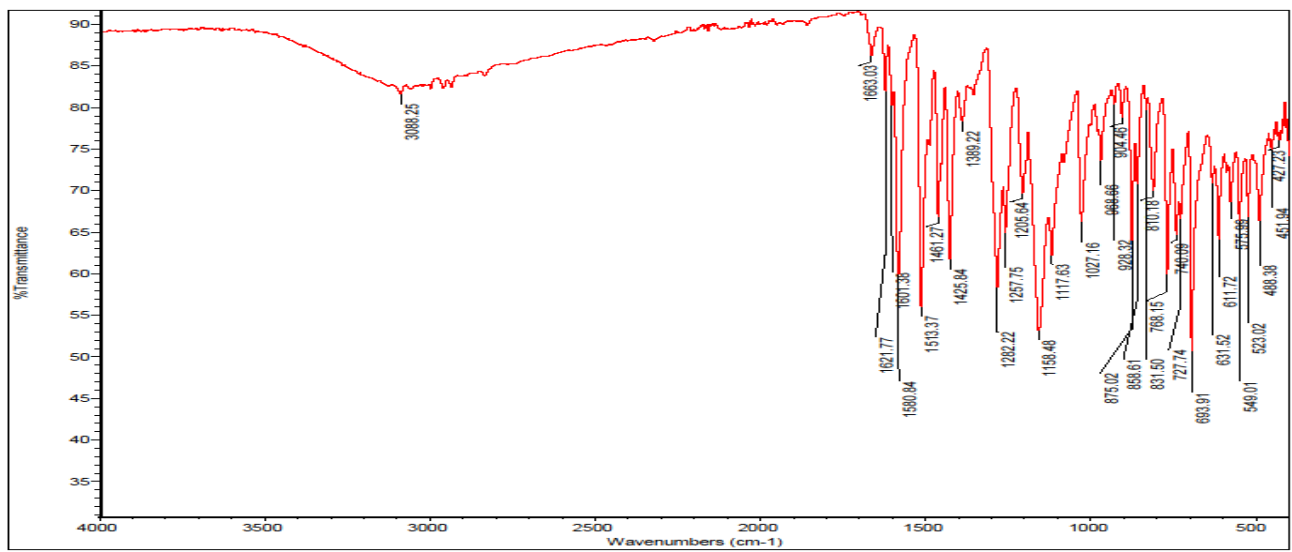

Fig. 5: FT-IR Spectra of Vanillin – Tryptophan Schiff base Ligand

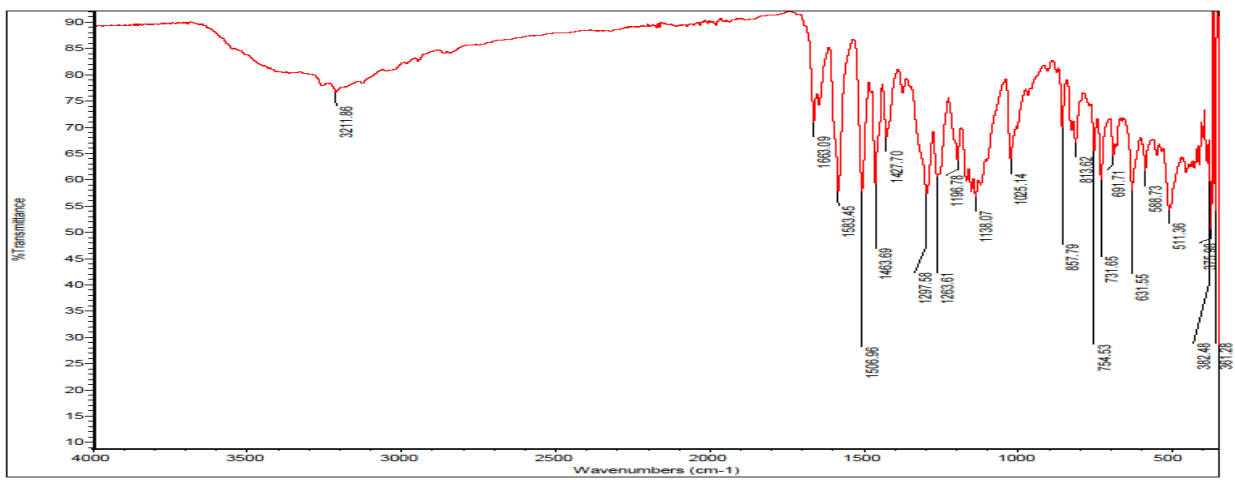

Fig. 6: FT-IR Spectra of Vanillin – Tryptophan – Cobalt (II) Complex

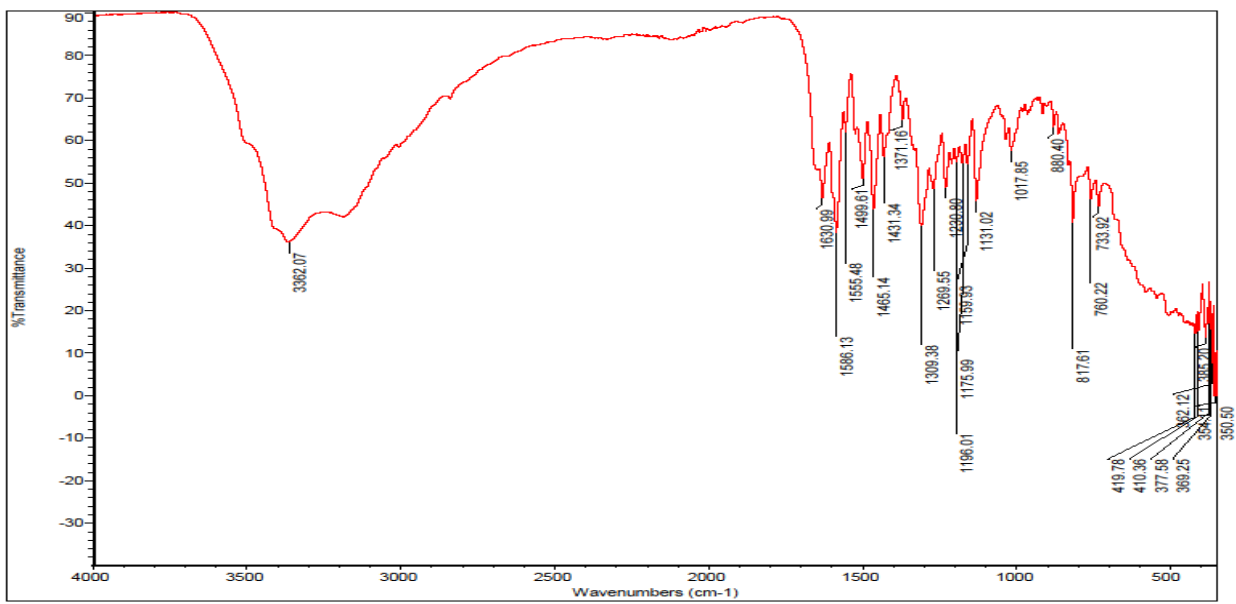

Fig. 7: FT-IR Spectra of Vanillin – Tryptophan – $Ni^{2+}$ Complex

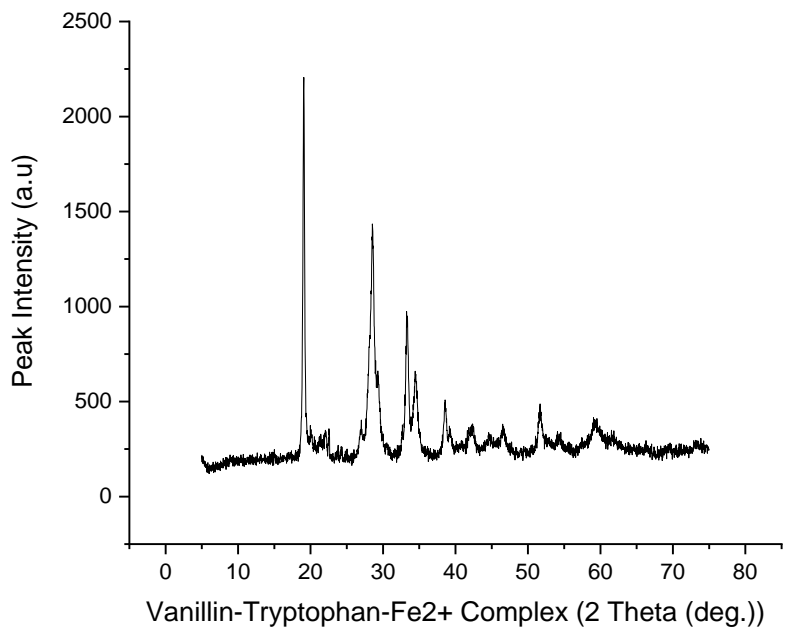

Fig. 8: PXRD Diffraction Pattern for Vanillin-Tryptophan-Fe2+ Complex

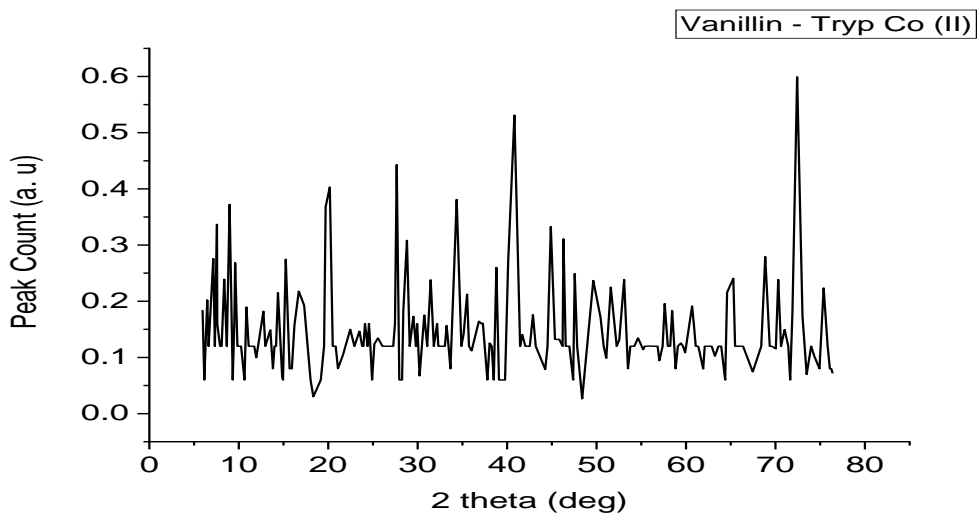

Fig. 9: Vanillin – Tryptophan - Co 2+ Complex

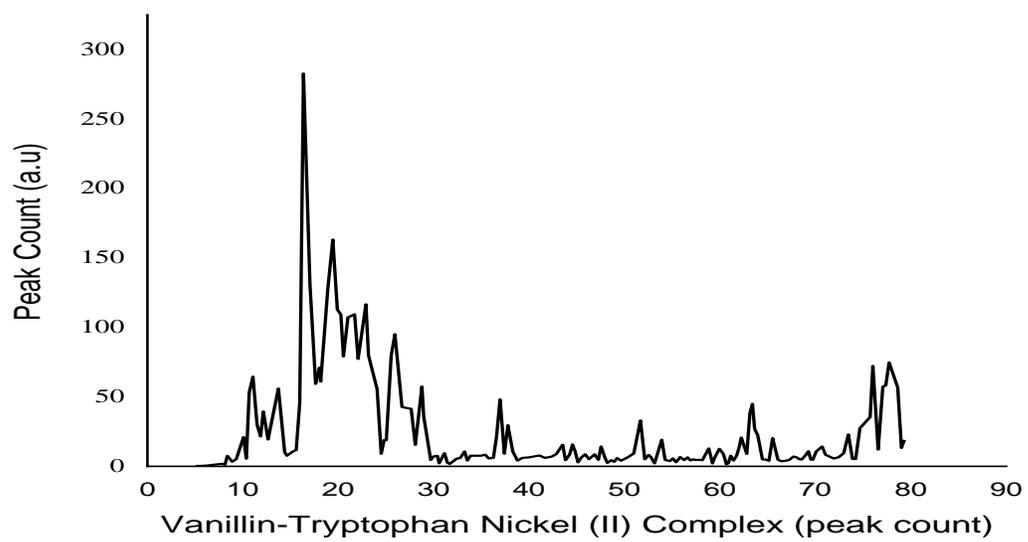

agent: a review. *American Journal of Heterocyclic Chemistry*, *4*(1), 1–21. https://doi.org/10.11648/j.ajhc.20180401.11

Kaczmarek, M. T., Jastrzab, R., Hołderna-Kedzia, E., and Radecka-Paryzek, W. (2009). Self-Assembled synthesis, characterization and antimicrobial activity of zinc(II) Salicylaldimine Complexes. *Inorganica Chimica Acta*, *362*(9), 3127–3133. https ://doi.org/10.1016/j.ica.2009.02.012

Maalik, A., Khan, F. A., Mumtaz, A., Mehmood, A., Azhar, S., Atif, M., Karim, S., Altaf, Y., and Tariq, I. (2014a). Within a Pharmacological Applications of Quercetin and its Derivatives: A short eview. *Tropical Journal of Pharmaceutical Research*, *13*(9), 1561–1566. https://doi.org/10.4314/tjpr.v13i9.26

Malik A, Goyat G, Vikas K, V. K. and G. S. (2018). *Coordination of Tellurium ( IV ) with Schiff Base Derived from o- Vanillin and*. *16*(Iv), 1–10.

M. S. Suresh and V. Prakash (2010). Preparation and characterization of Cr(III), Mn(II), Co(III), Ni(II), Cu(II), Zn(II) and Cd(II) chelates of schiffs base derived from vanillin and 4-Amino antipyrine *International Journal of the Physical Sciences* 5(14), pp. 2203-2211, Retrieved from at http://www.academicjournals.org/IJPS ISSN 1992 - 1950

Mustafa, S. K., and Alsharif, M. A. (2018). Copper ( Cu ) an Essential Redox-Active Transition Metal in Living System. *American Journal of Analytical Chemistry— A Review Article*. *0*, 15–26. https:// doi.org/10.4236/ajac.2018.91002.

Nagesh, G. Y., and Mruthyunjayaswamy, B. H. M. (2015). Synthesis, characterization and Biological relevance of some metal (II) complexes with oxygen, nitrogen and oxygen (ONO) donor Schiff base ligand derived from thiazole and 2-hydroxy-1-naphthaldehyde. *Journal of Molecular Structure*, *1085*(Ii), 198–206 https://doi.org/ 10.1016/ j.molstruc. 2014.12.058

Nathally Claudiane de Souza Santos, Regiane Bertin de Lima Scodro, Eloı́sa Gibin Sampiron, Andressa Lorena Ieque,2 Hayalla Correˆa de Carvalho, Thais da Silva Santos, Luciana Dias Ghiraldi Lopes, Paula Aline Zanetti Campanerut-Sá, Vera Lucia Dias Siqueira, Katiany Rizzieri Caleffi-Ferracioli, Jorge Juarez Vieira Teixeira and Rosilene Fressatti Cardoso (2020). Minimum Bactericidal Concentration Techniques in Mycobacterium tuberculosis: A Systematic Review.
25

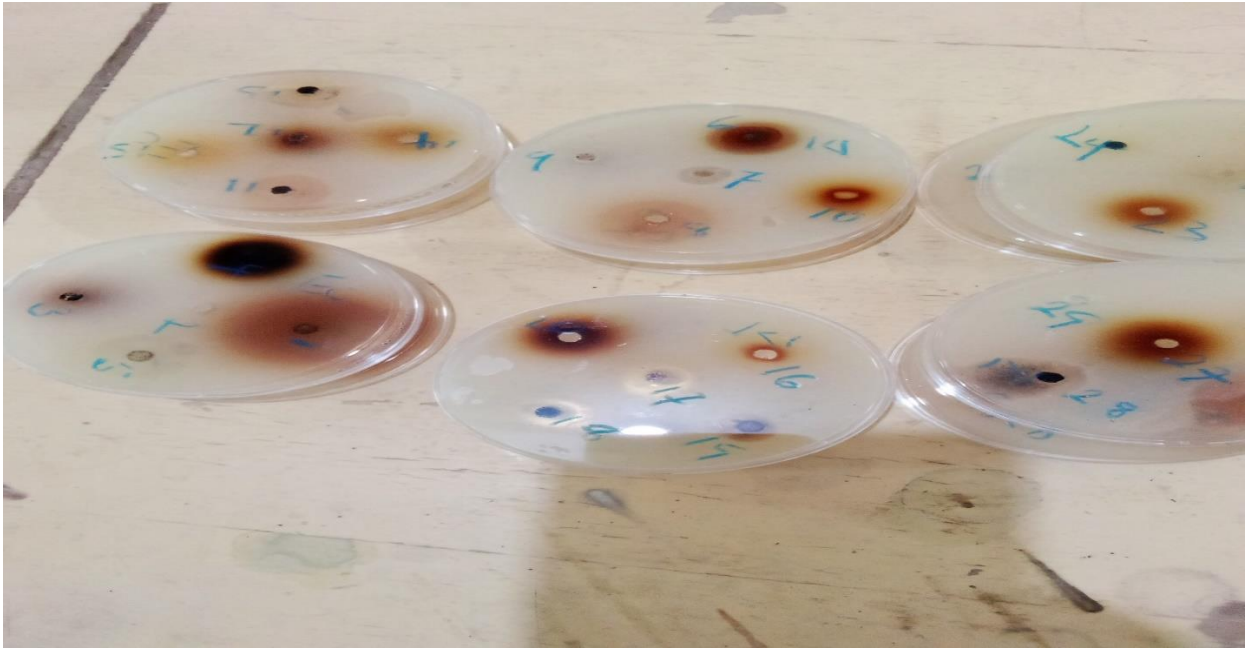
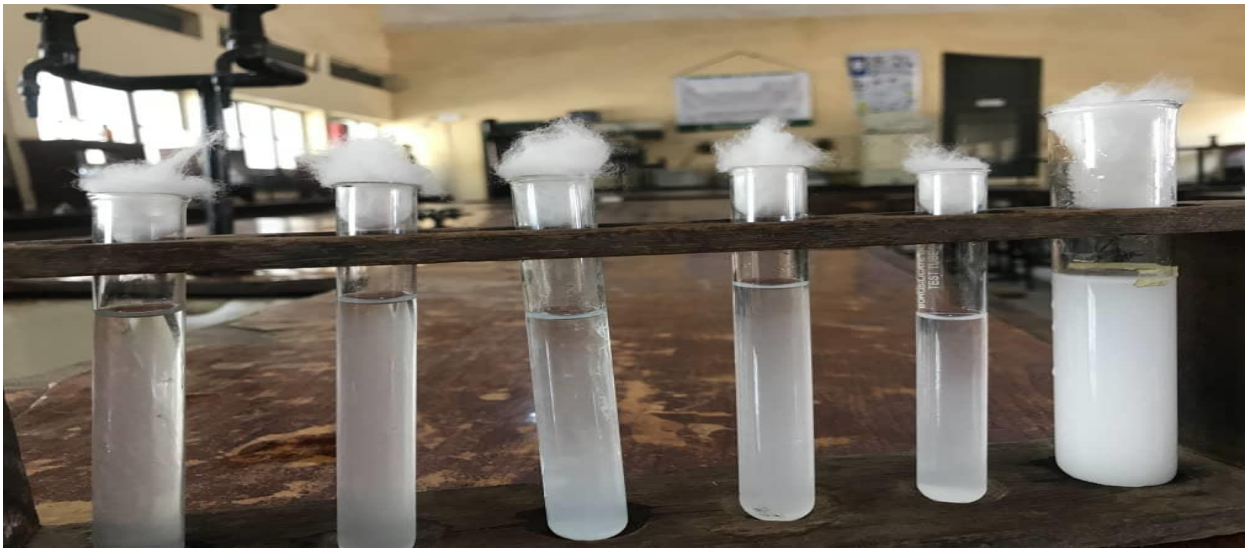

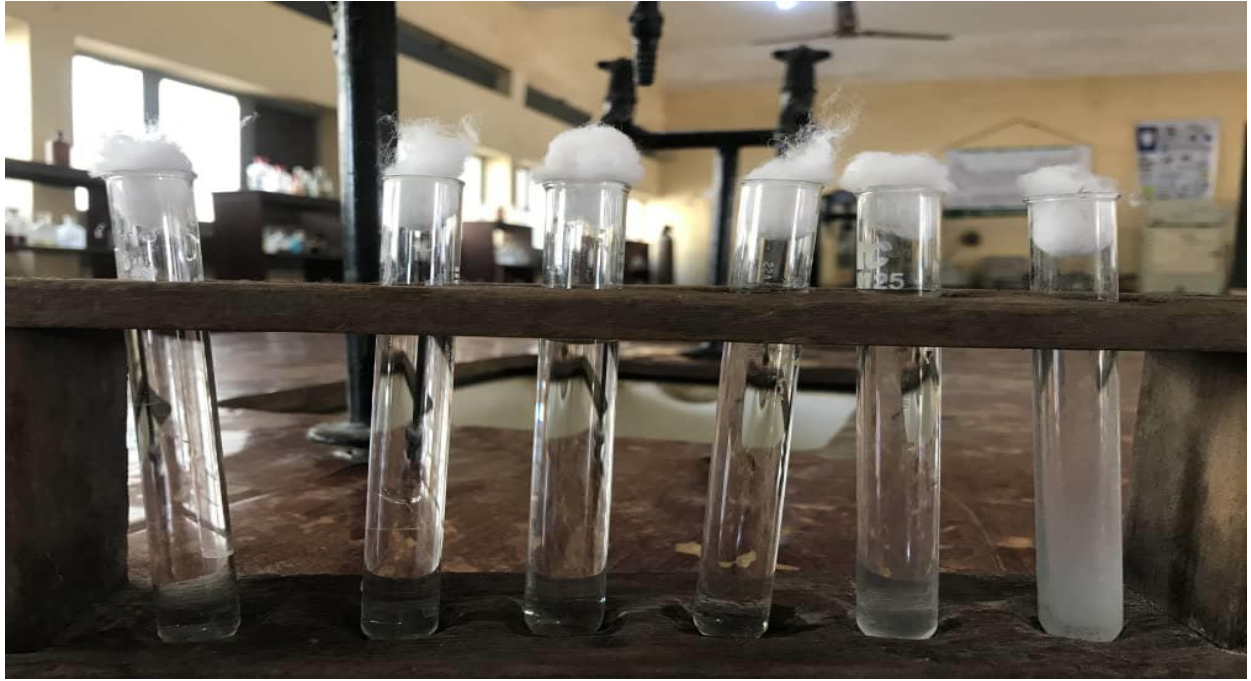